\documentclass[amsmath,amssymb]{article}

\usepackage{a4}
\usepackage{jheppub}

\usepackage{graphicx}
\usepackage{dcolumn}
\usepackage{bm}

\newcommand{\bdm}{\begin{displaymath}}
\newcommand{\edm}{\end{displaymath}}
\newcommand{\beq}{\begin{equation}}
\newcommand{\eeq}{\end{equation}}
\newcommand{\bea}{\begin{eqnarray}}
\newcommand{\eea}{\end{eqnarray}}

\newcommand{\be}{\begin{equation}}
\newcommand{\ee}{\end{equation}}

\begin{document}


\title{Addendum: Towards next-to-next-to-leading-log accuracy for the
  width difference in the $\mathbf{B_s-\bar{B}_s}$ system: fermionic
  contributions to order $\mathbf{(m_c/m_b)^0}$ and $\mathbf{(m_c/m_b)^1}$} 


\author[a]{Artyom~Hovhannisyan}%
\emailAdd{artyom@yerphi.am}
\affiliation[a]{%
  A. I. Alikhanyan National Science Laboratory (Yerevan Physics
  Institute), 0036 Yerevan, Armenia }

\author[b]{and Ulrich Nierste}%
\emailAdd{ulrich.nierste@kit.edu}%
\affiliation[b]{%
 Institut f{\"u}r Theoretische Teilchenphysik, Karlsruher
    Institut f{\"u}r Technologie, 76131 Karlsruhe, Germany
}


\abstract{%
  We calculate the three-loop master integrals of
  Ref.~\cite{Asatrian:2017qaz} in analytic form. This allows us to
  present the fermionic contributions to the $\Delta B=2$ Wilson
  coefficients of the $B$--$\bar B$ decay matrix in
  next-to-next-to-leading order of QCD with full analytic dependence on
  the mass of the charm quark in the fermionic loops.
}

\maketitle
\section{Introduction}
The off-diagonal element $\Gamma_{12}$ of the $2\times 2$ decay matrix
of the $B$--$\bar B$ mixing problem, where $B$ represents $B_s$ or $B_d$, must be
calculated to predict the width difference $\Delta \Gamma$ between the
$B$ meson mass eigenstates and the CP asymmetry in flavour-specific $B$
decays. The contributions with fermionic loops of the
next-to-next-to-leading order (NNLO) prediction calculated in
Ref.~\cite{Asatrian:2017qaz} are functions of $z=m_c^2/m_b^2$, where
$m_c$ and $m_b$ are the masses of charm and bottom quark,
respectively. The three-loop master integrals have been derived as an
expansion in $z$ and all but two integrals are given in analytic form,
while the remaining two integrals involve numerically calculated
coefficients.  In this Addendum, we present analytic results for these
as well and arrive at concise analytic results for the contributions of
current-current operators to the $\Delta B=2$
Wilson coefficients entering $\Gamma_{12}$. The corresponding
contributions with penguin operators are also known in
analytic form \cite{Asatrian:2020zxa}. 

\section{\bf Updated Master Integrals}
Our updated analytic results concern the imaginary parts
originating from four-particle-cuts of the master integrals in Eqs.~$(\text{B.2})$
and $(\text{B.6})$ of Ref.~\cite{Asatrian:2017qaz}. The  expansion in
$z=m_c^2/m_b^2$ up to $\mathcal O(z^4)$ in the $\overline{\rm MS}$
scheme reads:


\begin{eqnarray}
  \label{MI1} \nonumber && \hspace{-0.5cm} Im^{(4)} \int [dk] \frac{1}{\left(k_2^2-m_b^2\right)
                           k_3^2\left((k_1-p_b)^2-m_c^2\right)\left((k_1-k_2)^2-m_c^2\right)(k_2-k_3)^2}
  \\
                        &=& \nonumber \frac{(\mu/m_b)^{6 \epsilon-2 }}{8192 \pi ^5} \left[-\frac{7}{2}+\frac{\pi^2}{3} -z \left(4+\frac{2 \pi ^2}{3}+4 \log (z)\right)
                            + z^2 \left(\frac{27}{2}-7 \log (z)+\log ^2(z)\right)\right.
  \\
                        && \nonumber \hspace{1.3cm} + z^3 \left(-\frac{11}{3}+2 \log (z)\right) + z^4 \left(\frac{7}{24}+\frac{1}{2}\log (z)\right)
  \\
                        && \nonumber + \epsilon  \left(-\frac{175}{4}+\frac{3 \pi ^2}{2}+22 \zeta (3) +z \left(-56+\frac{2
                           \pi ^2}{3}-44 \zeta (3)-32 \log (z)+2 \log ^2(z)\right) + \frac{32\pi ^2}{3} z^{3/2}\right.
  \\
                        && \nonumber \hspace{0.7cm} +z^2 \left(-\frac{187}{4}+\frac{20 \pi ^2}{3}+12 \zeta (3)+\left(\frac{3}{2}-\frac{4 \pi ^2}{3}\right) \log
                           (z)+\frac{9}{2} \log ^2(z)-\log ^3(z)\right) - \frac{32 \pi ^2}{15} z^{5/2}
  \\
                        && \left.\left. \hspace{0.7cm} + z^3 \left(\frac{16}{3}-2 \pi^2-\frac{4}{3} \log (z)\right) - \frac{32 \pi ^2}{105} z^{7/2}
                           + z^4 \left(-\frac{6691}{1800}-\frac{\pi
                           ^2}{2}+\frac{43}{10}
                           \log(z)\right)\right)\right] \nonumber\\ 
        && \hspace{0.7cm}+ \mathcal{O}\left(z^5,\epsilon^2\right),
\end{eqnarray}
\begin{eqnarray}
\label{MI2} \nonumber && \hspace{-0.5cm} Im^{(4)} \int [dk] \frac{1}{k_1^2 (k_1-p_b)^2
\left((k_2-p_b)^2-m_b^2\right)(k_1-k_2)^2\left(k_3^2-m_c^2\right)\left((k_2-k_3)^2-m_c^2\right)}
\\
\nonumber &=&\frac{(\mu/m_b)^{6 \epsilon }}{8192 \pi ^5} \left[-4+\frac{\pi ^2}{3}+z \left(2-\frac{2 \pi ^2}{3}+12 \zeta (3)-\left(2-\frac{2 \pi ^2}{3}\right) \log (z)+\log^2(z)-\frac{1}{3}\log ^3(z)\right)\right.
\\
&& \left. \hspace{1cm} +z^2 \left(-\frac{19}{2}+3 \log (z)\right)+z^3
   \left(-\frac{1}{6}+\frac{1}{2}\log(z)\right)+z^4
   \left(\frac{67}{216}+\frac{1}{6}\log (z)\right)\right]
   \nonumber\\
  &&  \hspace{-0.5cm} + \mathcal{O}\left(z^5,\epsilon^1\right),
\end{eqnarray}
with the loop measure defined as
\begin{eqnarray}
\int [dk]=\int \frac{dk_1^d}{(2\pi)^d}\int
\frac{dk_2^d}{(2\pi)^d}\int \frac{dk_3^d}{(2\pi)^d}.
\end{eqnarray}
The Eq.~(\ref{MI1}) was obtained with the differential-equation method,
while for Eq.~(\ref{MI2}) we have used the formulas for the calculation
of four-particle phase space integrals derived in
\cite{Asatrian:2012tp}.

\section{\bf Analytic results for $\Delta B=2$ coefficients at order
  $\alpha^2_sN_f$ }
The results above entail the following updated analytic functions
$F_{ij}^{(2),N_v}(z)$ and $F_{S,ij}^{(2),N_v}(z)$ (determining the NNLO charm-loop 
contribution to the $\Delta B=2$ Wilson coefficients), superseding
Eqs.~(4.8)-(4.13) of Ref.~\cite{Asatrian:2017qaz}:
\begin{eqnarray}
\label{resz1} && \nonumber F^{(2),N_V}_{11}(z) = -\frac{386}{9} \log \left(\frac{\mu_1}{m_b}\right)+\frac{176}{9} \log\left(\frac{\mu_2}{m_b}\right)-\frac{40}{3} \log \left(\frac{\mu_1}{m_b}\right) \log \left(\frac{\mu_2}{m_b}\right)+\frac{20}{3} \log ^2\left(\frac{\mu_2}{m_b}\right)
\\
&&  \nonumber \hspace{2cm} -\frac{2689}{54}+\frac{5 \pi ^2}{9}+32 \zeta (3) - 4\pi^2\sqrt{z}+z \left(37-24\log (z)\right)-4\pi^2z^{3/2}
\\
&&  \nonumber \hspace{2cm} + z^2 \left(\frac{2219}{9}+\frac{8\pi^2}{9}-192\zeta (3)-\frac{572}{9} \log (z)+2\log^2 (z)\right)
\\
&& \hspace{2cm} +z^3 \left( \frac{46714}{2025}+\frac{128\pi^2}{27}+\frac{4846}{135} \log (z)-\frac{128}{9}\log ^2 (z)\right) + \mathcal O(z^4),
\end{eqnarray}
\begin{eqnarray}
\label{resz2} && \nonumber F^{(2),N_V}_{12}(z) = \frac{554}{27} \log\left(\frac{\mu_1}{m_b}\right)+\frac{352}{27} \log \left(\frac{\mu_2}{m_b}\right) - \frac{80}{9} \log\left(\frac{\mu_1}{m_b}\right) \log \left(\frac{\mu_2}{m_b}\right) + \frac{68}{3} \log^2\left(\frac{\mu_1}{m_b}\right)
\\
&&  \nonumber \hspace{2cm} + \frac{40}{9} \log^2\left(\frac{\mu_2}{m_b}\right) + \frac{3473}{324} + \frac{10 \pi ^2}{27} +
\frac{64 \zeta (3)}{3} - \frac{8\pi^2}{3} \sqrt{z}-z \left(\frac{334}{3} + 16 \log (z)\right)
\\
&& \nonumber \hspace{2cm}  - \frac{8\pi^2}{3} z^{3/2} + z^2 \left(\frac{7345}{27}-\frac{86 \pi ^2}{27} -128 \zeta (3) - \frac{3184}{27} \log (z) + \frac{55}{3} \log ^2 (z)\right)
\\
&& \hspace{2cm}  + z^3 \left(\frac{189512}{6075}+\frac{256\pi^2}{81} + \frac{8468}{405} \log (z) - \frac{256}{27} \log ^2 (z)\right) + \mathcal O(z^4),
\end{eqnarray}
\begin{eqnarray}
\label{resz3} && \nonumber F^{(2),N_V}_{22}(z) =  \left(\frac{236}{27}+\frac{4 \pi ^2}{9}\right) \log \left(\frac{\mu_1}{m_b}\right) + \frac{58}{27} \log \left(\frac{\mu_2}{m_b}\right) - \frac{32}{9} \log \left(\frac{\mu_1}{m_b}\right) \log \left(\frac{\mu_2}{m_b}\right) + \frac{20}{3} \log^2\left(\frac{\mu_1}{m_b}\right)
\\
&& \nonumber \hspace{2cm}  + \frac{16}{9} \log^2 \left(\frac{\mu_2}{m_b}\right) + \frac{3911}{324}+\frac{13 \pi ^2}{54} + \frac{16 \zeta (3)}{3} - \frac{16\pi^2}{3} \sqrt{z} - z \left(\frac{94}{3}-\frac{4 \pi ^2}{3}+32 \log (z)\right)
\\
&& \nonumber \hspace{2cm} + \frac{32\pi^2}{9} z^{3/2} + z^2 \left(\frac{70}{3}+\frac{40 \pi ^2}{27}-20 \zeta (3)-\frac{899}{54} \log (z)-\frac{17}{6} \log ^2(z)\right)
\\
&& \hspace{2cm} + z^3 \left(-\frac{28369}{6075}+\frac{40\pi^2}{81}+\frac{3764}{405} \log (z)-\frac{40}{27} \log ^2 (z)\right) + \mathcal O(z^4),
\end{eqnarray}
\begin{eqnarray}
\label{resz4} && \nonumber F^{(2),N_V}_{S,11}(z) = - \frac{80}{9} \log
   \left(\frac{\mu_1}{m_b}\right) + \frac{320}{9} \log \left(\frac{\mu_2}{m_b}\right) + \frac{128}{3} \log \left(\frac{\mu_1}{m_b}\right) \log \left(\frac{\mu_2}{m_b}\right) - \frac{64}{3} \log ^2\left(\frac{\mu_2}{m_b}\right)
\\
&& \nonumber \hspace{2cm} -\frac{470}{27}+\frac{56 \pi ^2}{9} + 32 \zeta (3) - 16 \pi^2\sqrt{z} +136 z - 16 \pi^2 z^{3/2}
\\
&& \nonumber \hspace{2cm} + z^2 \left(\frac{1964}{9}+\frac{80 \pi ^2}{9}-192 \zeta (3)-\frac{680}{9}\log (z)+8 \log ^2(z)\right)
\\
&& \hspace{2cm} + z^3 \left(\frac{43744}{2025}+\frac{128 \pi ^2}{27}+\frac{5296}{135} \log (z)-\frac{128}{9} \log ^2(z)\right) + \mathcal O(z^4),
\end{eqnarray}
\begin{eqnarray}
\label{resz5} && \nonumber F^{(2),N_V}_{S,12}(z) =  \frac{464}{27} \log \left(\frac{\mu_1}{m_b}\right)+\frac{640}{27} \log \left(\frac{\mu_2}{m_b}\right)
+ \frac{256}{9} \log \left(\frac{\mu_1}{m_b}\right) \log \left(\frac{\mu_2}{m_b}\right)+\frac{32}{3} \log^2\left(\frac{{\mu_1}}{m_b}\right)
\\
&& \nonumber \hspace{2cm} -\frac{128}{9} \log^2\left(\frac{\mu_2}{m_b}\right) + \frac{734}{81} + \frac{112 \pi ^2}{27} + \frac{64 \zeta (3)}{3} - \frac{32\pi^2}{3} \sqrt{z} + \frac{80}{3} z - \frac{32\pi^2}{3} z^{3/2}
\\
&& \nonumber \hspace{2cm} + z^2 \left(\frac{5296}{27}+\frac{112 \pi ^2}{27}-128 \zeta (3)-\frac{2320}{27} \log (z)+\frac{40}{3} \log ^2(z)\right)
\\
&& \hspace{2cm} + z^3 \left(\frac{132704}{6075}+\frac{256 \pi ^2}{81}+\frac{10016}{405} \log (z)-\frac{256}{27} \log ^2(z)\right) + \mathcal O(z^4),
\end{eqnarray}
\begin{eqnarray}
\label{resz6} && \nonumber F^{(2),N_V}_{S,22}(z) =
                 \left(\frac{704}{27}-\frac{32 \pi ^2}{9}\right) \log
                 \left(\frac{\mu_1}{m_b}\right) - \frac{320}{27} \log
                 \left(\frac{\mu_2}{m_b}\right) -\frac{128}{9} \log \left(\frac{\mu_1}{m_b}\right) \log \left(\frac{\mu_2}{m_b}\right)\\
  && \nonumber  \hspace{2cm} +\frac{32}{3} \log^2\left(\frac{\mu_1}{m_b}\right)
+\frac{64}{9} \log ^2\left(\frac{\mu_2}{m_b}\right) \\
&& \nonumber \hspace{2cm}   + \frac{2018}{81} - \frac{148 \pi ^2}{27}-\frac{32 \zeta (3)}{3}+ \frac{32\pi^2}{3} \sqrt{z} - z \left(\frac{208}{3}+\frac{32 \pi ^2}{3}\right) + \frac{608\pi^2}{9} z^{3/2}
\\
&& \nonumber \hspace{2cm}  + z^2 \left(-\frac{5552}{9}+\frac{352 \pi ^2}{27}-32 \zeta (3)+\frac{6380}{27} \log (z)-\frac{172}{3} \log ^2(z)\right)
\\
&& \hspace{2cm} + z^3 \left(-\frac{243592}{6075}+\frac{64 \pi ^2}{81}+\frac{11552}{405} \log (z)-\frac{64}{27} \log ^2(z)\right) + \mathcal O(z^4).
\end{eqnarray}
The contribution of any light quark $u,d,s$ can be obtained by setting
$z=0$ in eqs. (\ref{resz1})-(\ref{resz6}).

The new analytic results presented in this addendum are in excellent agreement with the previous numerical results presented in Ref.~\cite{Asatrian:2017qaz}. \\


{\it Acknowledgments}.  The work of A.H.\ has been supported by the
State Committee of Science of Armenia Program, Grant
No.~21AG-1C084. U.N.\ is supported by BMBF under grant
\emph{Verbundprojekt 05H2021 (ErUM-FSP T09) -- Belle II: Theoretische
  Studien f\"ur Belle II und LHCb} and by project C1b of the DFG-funded
Collaborative Research Center TRR 257, ``Particle Physics Phenomenology
after the Higgs Discovery''.

\end{document}